\documentstyle[12pt]{article}
\newenvironment{comment}[1]{}{}

\def\bib{\bibitem}
\def\be{\begin{equation}}
\def\ee{\end{equation}}
\def\beqar{\begin{eqnarray}}
\def\eeqar{\end{eqnarray}}
\def\barr{\begin{array}}
\def\earr{\end{array}}

\def\and{\qquad {\rm and } \qquad}
\def\etal{ {\it et al.} }
\def\ie{ {\it i.e.} }
\def\eg{ {\it e.g.} }

\def\kg{\kappa_\gamma}
\def\kZ{\kappa_Z}
\def\lg{\lambda_\gamma}
\def\lZ{\lambda_Z}

\def\gsw{$SU(2)_L \otimes U(1)_Y$ }
\def\prl#1{Phys. Rev. Lett. {\bf #1}}
\def\prd#1{Phys. Rev. {\bf D#1}}
\def\plb#1{Phys. Lett. {\bf B#1}}
\def\npb#1{Nucl. Phys. {\bf B#1}}

\begin{document}
\thispagestyle{empty}
\setcounter{page}{0}
\renewcommand{\thefootnote}{\fnsymbol{footnote}}

\begin{flushright}
MPI-Ph/93-98\\
December 1993
\end{flushright}

\vspace{5ex}
\begin{center}

{\Large \bf Testing Anomalous $W$ Couplings in $e^- e^-$ Collisions}\\

\bigskip
\bigskip
{\sc
   Debajyoti
Choudhury\footnote{debchou@iws166.mppmu.mpg.de,debchou@dmumpiwh.bitnet}
   and
   Frank Cuypers\footnote{cuypers@iws166.mppmu.mpg.de}
   }

\bigskip
{\it Max-Planck-Institut f\"ur Physik, Werner-Heisenberg-Institut, \\
F\"ohringer Ring 6, \\
80805 M\"unchen, Germany.}

\bigskip
\bigskip
{\bf Abstract}
\end{center}

We analyze the influence of anomalous gauge couplings
in the reaction
$e^- \: e^- \longrightarrow e^- \: W^- \: \nu_e$
at a 500 GeV linear collider.
The limits imposed by this  process
on deviations from the standard model of electro-weak interaction,
are competitive with those inferred from
other high energy experiments. Furthermore, the allowed domain in the
parameter space is quite different, and hence such an experiment would
more than complement the other direct searches.

\newpage
\setcounter{footnote}{0}
\renewcommand{\thefootnote}{\arabic{footnote}}

The success of the LEP experiments in measuring the $Z$-boson mass and its
couplings to fermions to an accuracy level of better than $1\%$ has led to
a dramatic confirmation of the predictions of the
\gsw  theory of electroweak interactions as embedded in the standard model
(SM). A measure of this success is reflected in the recent effort to further
probe the SM in the gauge sector so as to be able to experimentally establish
it as ruled by a gauge principle. Indeed, some progress has already been
achieved through the observation of the process $p\bar{p} \rightarrow e
\nu \gamma X$ at the Collider Detector at Fermilab (CDF) \cite{CDF} and at
UA2 at CERN \cite{UA2}.
Interpreting this to be a signal of  $W \gamma$ production and subsequent
decay of the $W$, limits were put on possible deviations of the $WW\gamma$
vertex from its SM structure \cite{rahul-1}. These limits are very weak
though, especially when compared with the contributions from the
one-loop corrections in the SM
\cite{couture}.
Recently,
this has led to a lot of work
in identifying better signals for such deviations especially in the context of
LEP-200,
LHC and SSC \cite{LEP2,kane} as well as as HERA \cite{Snowmass}.
Several studies have also been performed for the case of linear $e^+ e^-$
colliders \cite{rahul-nita,photon}
as well as for $e \gamma$ and $\gamma \gamma$
colliders \cite{photon},
where a high energy photon beam is obtained
by back-scattering an intense laser ray
off a electron beam at a linear collider.

In this Letter,
we advocate the use of electron-electron collisions
to probe the gauge sector of the SM.
Such beams can in principle be easily obtained at a linear collider
of the next generation
(CLIC, JLC, NLC, TESLA, VLEPP, \dots).
We concentrate here on a machine operating at 500 GeV
and able to accumulate 1 to 10 fb$^{-1}$ of integrated luminosity.

In principle,
any deviation of a gauge coupling from its SM value
would contribute significantly in the estimation
of quantum corrections.
This argument was used to constrain these couplings
using LEP data as well as other low energy measurements
\cite{couture,kane,derujula}.
However, it was pointed out \cite{burgess} that most of these
calculations made an improper use of the cut-off procedure and as a result had
grossly overestimated such constraints. Subsequently, a number of studies were
made both in the context of a decoupling Lagrangian \cite{zeppen} as well as
the case of a non-linearly realized \gsw symmetry with an unspecified Higgs
sector \cite{crs}, to obtain various constraints. All such efforts suffer
though from one of two pitfalls:
either a model dependence or crucial
assumptions about the relative magnitudes of various effects, and hence are no
substitute for direct measurements.

The most general triple electroweak vector boson (TEVB)
coupling can be parametrised in the form of an
effective Lagrangian with seven parameters for each of the neutral
vector bosons \cite{peccei}.
We shall be more restrictive, though. Since the upper
limit on the neutron electric dipole moment restrict (barring large
cancellations between different contributions) the $CP$ violating
parameters to be less than $ O(10^{-4})$, we choose to neglect them altogether.
Furthermore, we demand individual $C$ and $P$ invariance for the TEVB
vertices. The Lagrangian for the $WWV \:(V=\gamma/Z)$ vertex
can then be expressed as
\be
     {\cal L}_{\it eff}^{V}= -i g_{V}
             \left[ g_1^V
               \left( W^\dagger_{\alpha \beta} W^\alpha
                      - W^{\dagger\alpha} W_{\alpha \beta}
                \right) V^\beta
             +
               \kappa_V  W^\dagger_{\alpha} W_\beta
                               V^{\alpha\beta}
            + \frac{\lambda_V}{M_W^2}
                 W^\dagger_{\alpha \beta} {W^\beta}_\sigma
                 V^{\sigma\alpha} \right]
      \label{lagrangian}
\ee
where $V_{\alpha\beta} = \partial_\alpha V_\beta - \partial_\beta
V_\alpha $ and  $W_{\alpha\beta} = \partial_\alpha W_\beta -
\partial_\beta W_\alpha $.  In (\ref{lagrangian}),
 $g_{V}$ measures  the $WWV$ coupling
strength in the SM with $g_{\gamma}=e$ and $g_{Z} = e \cot \theta_W$.
Whereas electromagnetic gauge invariance forces $g_1^\gamma =1 $, the
other couplings are model dependent and the tree level SM values are
$g_1^Z= \kg =\kZ =1$ and $\lg=\lZ=0$.

\begin{comment}{
The $WW\gamma$  parameters
lead to non-trivial  static magnetic dipole and electric
quadrupole moments $\mu_{W}$ and $Q_{W}$, respectively, of the W boson
\be
\barr{rcl}
\mu_{W}&=&\displaystyle\frac{e}{2M_{W}}(1+\kappa_{\gamma}+\lambda_{\gamma})
        \\[1.5ex]
Q_{W}&=&\displaystyle \frac{e}{2M_{W}}(\lambda_{\gamma}-\kappa_{\gamma})
\earr
\ee
}\end{comment}

Much criticism has been levelled at the lagrangian (\ref{lagrangian})
for its apparent lack of gauge invariance.
It has been shown recently \cite{burgess},
however, that any Lorentz and $U (1)_{\rm em}$  gauge invariant
Lagrangian, containing $W's$ and  $Z's$, automatically obeys
\gsw gauge invariance, realized nonlinearly in general.
Though the deviations of the above parameters from their SM values, as
viewed in the context of effective field theories, are expected to be tiny,
in some scenarios they can be very large indeed \cite{rizzo}.
We shall not comment on the
possible source of such deviations, but rather concentrate on their
measurability. Note that the use of (\ref{lagrangian}), rather than
the equivalent full chiral lagrangian \cite{chiral} is well-justified in such
a phenomenological study.

Having established the formalism, we now turn to the laboratory.
The process we consider here is
\be
e^- \: e^- \longrightarrow e^- \: W^- \: \nu_e
\ .
\label{ee2enw}
\ee
The final state is thus exceedingly simple:
an electron, a $W^-$ boson and missing momentum.
Note that unlike in most signals in hadronic or $e^+ e^-$ collisions,
there is no  $s$-channel process here.
At tree level the contributing diagrams,
shown in Fig.~1,
are of two types :
(i) a $t$ or $u$-channel
exchange of a vector boson with a
``$W$-Bremsstrahlung'' off one of the four fermions,
or (ii) those involving a trilinear coupling.
Only the diagrams pertaining to the second class
are of course sensitive to anomalous couplings.

The analytical expressions for the matrix elements were obtained
as a function of the five anomalous parameters
$g_1^Z,\ \kg,\ \kZ,\ \lg$ and $\lZ$,
with the aid of the software system
COMPHEP \cite{COMPHEP} and by an independent helicity-amplitude
calculation.
The Monte-Carlo routine VEGAS \cite{VEGAS}
was used for the numerical integration over phase space.

For the electron to be visible,
we required that its transverse momentum be greater than 5 GeV
and that its absolute rapidity should not exceed 3.
These requirements also eliminate a very large fraction of the
contributions originating from the $W$-Bremsstrahlung diagrams alone,
thus enriching the signal to background ratio.
In addition,
the rapidity cut ensures that the internal photon
never comes close to mass shell
and hence provides an easily integrable matrix element squared.
Because the $W$ boson undergoes a further decay
we have imposed no kinematical cut on its momentum. Explicit computations
show though that similar cuts on the $W$--momentum
have only a marginal effect.
Since a $e^- e^-$ machine
suffers very little background hadronic activity,
the reconstruction of a single $W$
should thus be straightforward and we shall assume, in this study, an
efficiency of 100\%.
In the SM limit
the total cross-section amounts to $\sigma_{SM}=1.49$ pb \cite{cor}.

An effort to deal simultanously
with all five parameters in of the lagrangian (\ref{lagrangian})
is bound to lead to a great deal of confusion.
We therefore first present in Fig.~2
the dependence of the difference between the SM and anomalous cross-sections
on each of the five parameters
with the others assuming their tree-level SM values.
This yields some idea about the detectability of the TEVB
couplings at such a collider.
In addition it offers a check of our calculations in that the
dependence on $\kappa_\gamma$ and $\lambda_\gamma$ is similar to
that obtained in
ref.~\cite{Snowmass} for the analogous process
$e p \rightarrow e W^\pm \:{\rm jets}$ at HERA.

The dependence on the $WW\gamma$ couplings is much more
pronounced than that on the $WWZ$ coupling.
This is easily traced back to the
ratio of the $\gamma$- and the $Z$-couplings
both to the fermions and the $W$ and
is in marked contrast to the case of their contributions to
the precison electroweak parameters \cite{zeppen,crs}.
The different sensitivity there
arises from the fact that the most accurate electroweak measurements are
those dealing with the properties of the $Z$.
Furthermore, the dependence on
$\lambda_{\gamma,Z}$ is weaker than that on $\kappa_{\gamma,Z}$.
However,
since $\lambda_{\gamma,Z}$ represent interaction terms of dimension 6 or
higher,
the sensitivity to these parameters
grows faster with the interaction energy  and
hence they would be more easily detectable at machines operating at a  higher
energy, {\it e.g.} the LHC.

To assess the resolving power of the process (\ref{ee2enw}),
we require the absolute difference between the SM and anomalous cross-sections
$\Delta\sigma = |\sigma_{\rm SM}-\sigma_{\rm anom}|$
to be large enough to provide a deviation from the SM prediction
exceeding the Poisson fluctuations by $N$ standard deviations \ie, the event
numbers ($n = {\rm luminosity} \times \sigma$) should satisfy the relation
$\Delta n > N \sqrt{n_{\rm SM}}$.
The required luminosity $\cal L$ to reach this goal
is then
\be
{\cal L} > {N^2\sigma_{SM} \over \Delta\sigma^2}
\ .
\label{lum}
\ee
To compare the resolving power of this reaction with the projections
for  conventional linear colliders,
we present the 90\% C.L. (N = 1.64 in Eq.~(\ref{lum})) limits to which it can
constrain these parameters for an
integrated luminosity of $10 \:{\rm fb}^{-1}$. Under the similar assumption
that only one of the parameters may be non--zero, these are then
\be
\barr{rclcrcl}
 \kappa_\gamma & : & (0.982,1.018  )
        & \qquad & \kappa_Z & : & (0.89,1.09)  \\
 \lambda_\gamma & : & (-0.08,0.09)  & \qquad & \lambda_Z & : & (-0.13,0.15)\\
                &   &         & \qquad & g^1_Z & : & (0.90,1.075)
\earr
\ee
These results are comparable
to those obtained in Refs~\cite{kane,rahul-nita,photon}.
The slightly tighter constraints
these authors obtain for some of the parameters,
can be traced back to a more detailed analysis
of the angular distributions,
which goes beyond the scope of this Letter.
Without such an analysis
(\eg Kalyniak \etal in \cite{rahul-nita}),
the limits are weaker than what we achieve from a
straightforward comparison of the total cross sections.

Although this analysis provides us with grounds for optimism, yet
some circumspection is called for.
Indeed,
it is rather unlikely that,
if at all,
then only one parameter assume an anomalous value
while all other remain at their SM values.
Even if one were to neglect the effects
of interference between the different contributions on grounds of their
representing operators of higher dimension (though  such an
argument is invalid in a phenomenological analysis such as the present), we
still would have to worry about incoherent addition.
The best approach then is to obtain contour plots.
As we are still plagued by the plethora of parameters,
we choose to present only two combinations
(in each of which the other parameters are assumed to have their SM values).

Assuming only $\kappa_\gamma -1 $ and $\kappa_Z - 1$ to be non-zero,
Fig.~3 shows the $3 \sigma$ band
(according to Eq.~(\ref{lum}), with $N=3$)
to which these parameters
can be restricted to with a total integrated luminosity of 1 fb$^{-1}$.
Clearly,
the interference between the $\kappa_\gamma$ and the $\kappa_Z$
contributions can be large,
thus rendering even a large value of either unobservable.
An increase in luminosity would not change this fact in any way,
for we would still have a band of the same average
dimensions, albeit narrower.
One may argue however, that from a theoretical standpoint, such
large cancellation are unlikely. We have hence concentrated more on the
region around the SM point (Fig.~4) and show the reachability
for a integrated luminosity of both 1 and 10 fb$^{-1}$.
In Fig.~5, we show similar contours for the
$\lambda_\gamma - \lambda_Z$ pair of parameters.
Unlike the $\kappa$-case,
here we do not observe a ``mexican hat'' band
but a simple well centered around the SM values 0.
As expected, the sensitivity to the $\lambda$'s
is much weaker than that to the $\kappa$'s.

To conclude,
we demonstrate the considerable power of a $e^- e^-$ collider
as a tool for  unravelling the self-interactions of the electroweak gauge
bosons. In fact, the limits that can be obtained from a simple minded analysis
of the total cross-sections alone compares quite favourably with those
deduced from a sophisticated analysis of the decay distributions in a
more conventional collider. However, what accords even greater value to such
an experiment is the significant difference in the region of the parameter
space that it would probe. We believe, hence, that it would prove to be of
great import when used in conjunction with the other experiments.

\bigskip
\bigskip
\bigskip

We are very much indebted to Edward Boos and Michael Dubinin
for having provided us with the COMPHEP software.
Many thanks also go to Geert Jan van Oldenborgh
for sharing with us his phase space integration routines.

\newpage
\centerline{\bf Figure Captions}
\begin{enumerate}
\item Lowest order Feynman diagrams
contributing to the process in Eq.~(\protect\ref{ee2enw}).
\item Difference in cross-sections
of the SM and anomalous processes of Eq.~(\protect\ref{ee2enw})
as a function of each of the five
anomalous couplings in Eq.~(\protect\ref{lagrangian}),
when all others assume their SM values.
The SM cross-section amounts to 1.49 pb.
\item Contours of detectability at the $3\sigma$ level
($N=3$ in Eq.~(\protect\ref{lum}))
of the $\kappa_\gamma$ and $\kappa_Z$ parameters
for an integrated luminosity of 1 fb$^{-1}$.
All other parameters assume their SM value.
\item Blowup of Fig.~3
in the neighbourhood of the SM values.
The contours for 10 fb$^{-1}$ are also shown.
\item Same as Fig.~4
for the $\lambda_\gamma$ and $\lambda_Z$ parameters.
\end{enumerate}
\end{document}